\def\ifmath#1{\relax\ifmmode #1\else $#1$\fi}%
\def\rB{\ifmath{{\mathrm{B}}}}
\def\rd{\ifmath{{\mathrm{d}}}}
\def\rD{\ifmath{{\mathrm{D}}}}
\def\rK{\ifmath{{\mathrm{K}}}}
\def\rp{\ifmath{{\mathrm{p}}}}
\def\rP{\ifmath{{\mathrm{P}}}}

\documentclass{ws-p8-50x6-00}

\begin{document}

\title{Self-Organized Criticality in Gluon Systems and its Consequences}

\author{K. Tabelow}

\address{Institut f\"ur Theoretische Physik, FU Berlin, Arnimallee 14, 14195
Berlin,Germany\\E-mail: karsten.tabelow@physik.fu-berlin.de}


\maketitle

\abstracts{It is pointed out, that color-singlet gluon clusters can be
  formed in hadrons as a consequence of self-organized criticality
  (SOC) in systems of interacting soft gluons, and that the properties
  of such spatiotemporal complexities can be probed experimentally by
  examing inelastic diffractive scattering. Theoretical arguments and
  experimental evidence supporting the proposed picture are presented.
  As a consequence of the space-time properties of the color-singlet
  gluon clusters due to SOC in gluon systems, a simple analytical
  formula for the differential cross section for inelastic diffractive
  hadron-hadron scattering can be derived. The obtained results are in
  good agreement with the existing data. Furthermore, it is shown that
  meson radii can be extracted from inelastic diffractive scattering
  experiments. The results obtained for pion and kaon charge radii are
  compared with those determined in meson form factor measurements.}

This talk is a short summary of three
papers\cite{letter,longSOC,meson} written at FU Berlin in
collaboration with C. Boros, T. Meng, R. Rittel, and Y. Zhang.

In 1987, Bak, Tang and Wiesenfeld (BTW) observed\cite{BTWoriginal} that
open, dynamical, complex systems far from equilibrium may evolve into
critical states, where local perturbations can propagate like
avalanches (called BTW-avalanches) over all length and time scales.
In contrast to critical behavior in equilibrium thermodynamics, the
evolution into this very delicate state needs {\em no fine tuning} of
external parameters: The criticality is self-organized (SOC). The
distributions of the size $S$ and the lifetime $T$ of such
BTW-avalanches obey power laws:
\begin{equation}
\label{fingerprints}
D_S(S)\propto S^{-\mu} \,\, \mbox{ and } \,\, D_T(T)\propto T^{-\nu}
\end{equation}
known as the ``fingerprints of SOC''.\cite{BTWcontinue} Since the
first observation of SOC, there has been vast interest in this fast
developing field. The reason is that SOC provides so far the only
known mechanism to generate spatiotemporal complexity, which is
ubiquitous in Nature. In the macroscopic world, many complex systems 
have been found\cite{BTWcontinue} showing this kind of behavior: sand-
and rice-piles, the crust of earth exhibiting earthquakes of all sizes,
the stock market and even the system of biological genera on earth.
As particle physicists we ask the following questions:
\begin{itemize}
\item Does SOC also exists at the fundamental level of matter, in the
  world of quarks and gluons?
\item How to probe this experimentally? What are the measurable
  consequences?
\end{itemize}

At the beginning, we search for an appropriate candidate for a system
to investigate and recall experimental results and theoretical facts
that are of considerable importance: First, it has been
observed\cite{HERAreview} that soft gluons dominate the small-$x_\rB$
region of deep-inelastic electron-proton scattering (DIS) at HERA.
Thus, there are many soft gluons in the proton. Second, gluons may
{\em directly} interact with each other through local gluon-gluon
coupling prescribed by the QCD-Lagrangian. This distinguishes their
behavior from that of other elementary particles such as photons.
Third, due to emission and absorption of gluons, the number of gluons
is not a conserved quantity.  This shows, that systems of interacting
soft gluons should be considered as {\em open, dynamical, complex
  systems with many degrees of freedom}. This can be considered as the
zeroth fingerprint for the existence of SOC in such systems.

What are the possible BTW-avalanches in the gluon system in the
proton? Another experimental observation is very helpful in this
connection: Large-rapidity gap events (LRG) have been
observed\cite{HERAreview} in the small-$x_\rB$ region of DIS, the same
region where the soft gluons dominate. In such events, the virtual
photons encounter colorless objects originating from the proton
(called Pomeron in Regge phenomenology\cite{HERAreview}). Therefore,
this type of events is very similar to inelastic diffractive
hadron-hadron scattering in which such colorless objects also play the
dominating role. The interaction of these color-singlets with the
proton remnant should be of Van-der-Waals type. Thus, they can be
easily knocked out by the projectile in reactions with relatively
small momentum transfer. This makes it relatively easy to ``examine''
them without the proton remnant in inelastic diffractive scattering
processes.

\begin{figure}[t]
\epsfxsize=24pc 
\centerline{\epsfbox{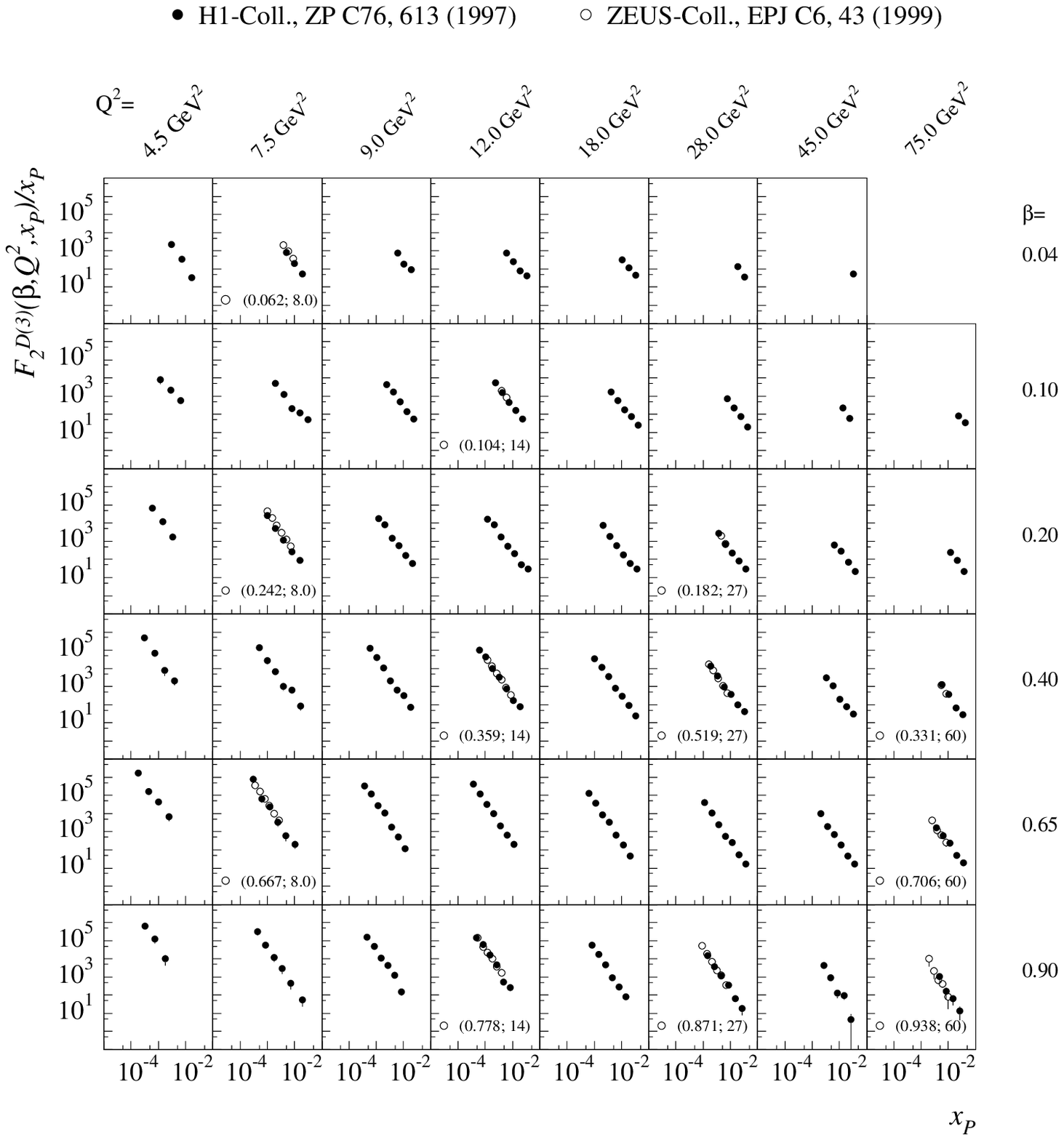}} 
\caption{$F_2^{\rD(3)}(\beta ,Q^2,x_\rP)/x_\rP$ as a function of $x_\rP$ 
  for a wide range of the kinematical variables $Q^2$ and $\beta$ in a
  double-logarithmic plot. This figure shows, that the power-law
  behavior with an exponent $\approx -2$ is independent of those
  variables.  Data are taken from Ref.[\ref{f2d3h1},\ref{f2d3zeus}].
\label{fig:fingerprint2}}
\end{figure}

It has been suggested,\cite{letter,longSOC} that the colorless
objects are BTW-avalanches in form of color-singlet gluon clusters due
to SOC in the system of interacting soft gluons in the proton. In
order to test this proposal, a systematic analysis of the existing
data has been performed,\cite{longSOC,rittel} where we made use of the
fact, that the size of the cluster is the measurable quantity $x_\rP$
and that the diffractive structure function $F_2^{\rD(3)}$ divided by
$x_\rP$ can be interpreted as the size distribution of the clusters. The
result of the analysis shows (see fig. \ref{fig:fingerprint2}), that
the size and the lifetime distribution\cite{longSOC} of the
color-singlet gluon clusters indeed show power-law behavior
Eq.(\ref{fingerprints}) with exponents $\mu \approx 2$ and $\nu
\approx 2$. Furthermore, this behavior is universal and can be seen in
all inelastic diffractive scattering processes (for details see
Ref.[\ref{longSOC}]), as it is expected if SOC is the underlying
production mechanism for the colorless objects.

The following physical picture for inelastic diffractive scattering
processes emerges: The beam particle sees a cloud of color-singlet
gluon clusters due to self-organized criticality (SOC) in the gluon
system. The size and lifetime distribution of the clusters obey
universal power laws. There is no typical size, no typical lifetime
and no static structure for the clusters. The color-singlet gluon
cluster is knocked out by the beam particle. Both break up producing
the usually unidentified hadronic system $X$.

\begin{figure}[t]
\epsfxsize=14pc 
\centerline{\epsfbox{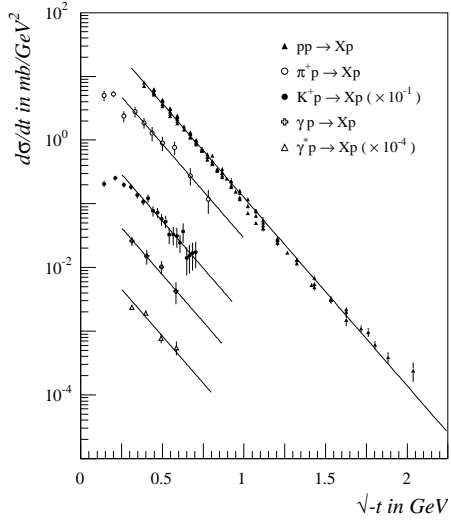}} 
\caption{The differential cross section $\rd\sigma /\rd t$ for inelastic 
  diffractive scattering off proton using different projectiles (p,
  $\pi^+$, $\rK^+$, $\gamma$ and $\gamma^\star$) as function of
  $\sqrt{-t}$. The data are taken from
  Refs.[\ref{albrow}-\ref{breitweg2}]. The solid lines represent the
  $\exp (-2a\sqrt{-t})$-dependence, where $a=\sqrt{\frac{3}{5}}r_\rp$
  with the proton radius $r_\rp$. See text or
  Refs.[\ref{letter},\ref{longSOC}] for details.
\label{fig:proton}}
\end{figure}

What are the measurable consequences?  Since the space-time
properties of the cluster cloud are known, we choose an optical
geometrical approach, where the beam particle is considered as high
frequency wave passing the target cloud of colorless gluon clusters.
Taking into account 1. the properties of the gluon clusters due to
SOC, 2. the confinement of single (colored) gluons, 3.  causality and
4. the cluster distribution, it is possible to
determine\cite{letter,longSOC,rittel} the differential cross section
for inelastic diffractive scattering processes
\begin{equation}
\label{crosssection}
\frac{\rd\sigma}{\rd t} = C \cdot \exp (-2 a \sqrt{-t}),
\end{equation}
where $t$ is the square of the four-momentum transfer and
$a=\sqrt{\frac{3}{5}}r_\rp$ is directly related to the proton radius and
$C$ is an unknown normalization constant. This behavior is an until
now unknown regularity and in very good agreement with the
experimental data for inelastic diffractive proton-proton and
(antiproton-proton) scattering.\cite{letter}

Here, we present a further test\cite{tabelow} by comparing
Eq.(\ref{crosssection}) with the experimental data for inelastic
diffractive scattering off protons using {\em different} projectiles,
e.g.
\begin{displaymath}
\rp, \gamma , \gamma^\star, \rK^+ , \pi^+
\end{displaymath}
In fig. \ref{fig:proton} it is shown, that the behavior of the 
differential cross section $\rd\sigma /\rd t$ is indeed given by 
Eq.(\ref{crosssection}), independent of the projectile. 

\begin{figure}[t]
\epsfxsize=14pc 
\epsfbox{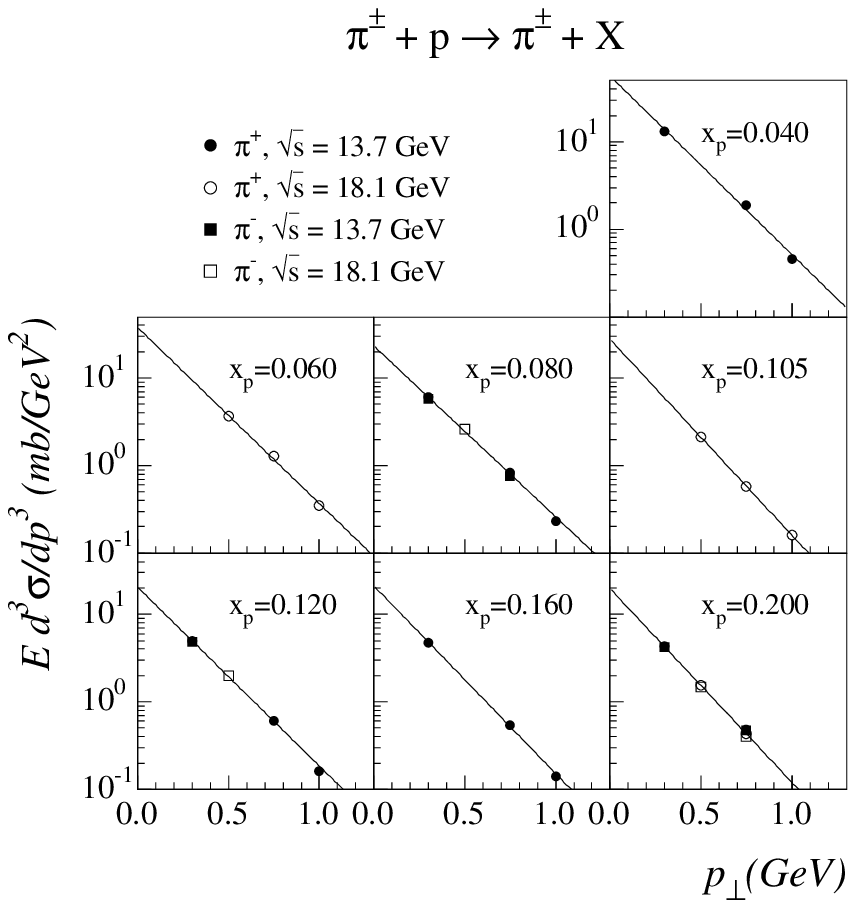} 
\epsfxsize=14pc 
\epsfbox{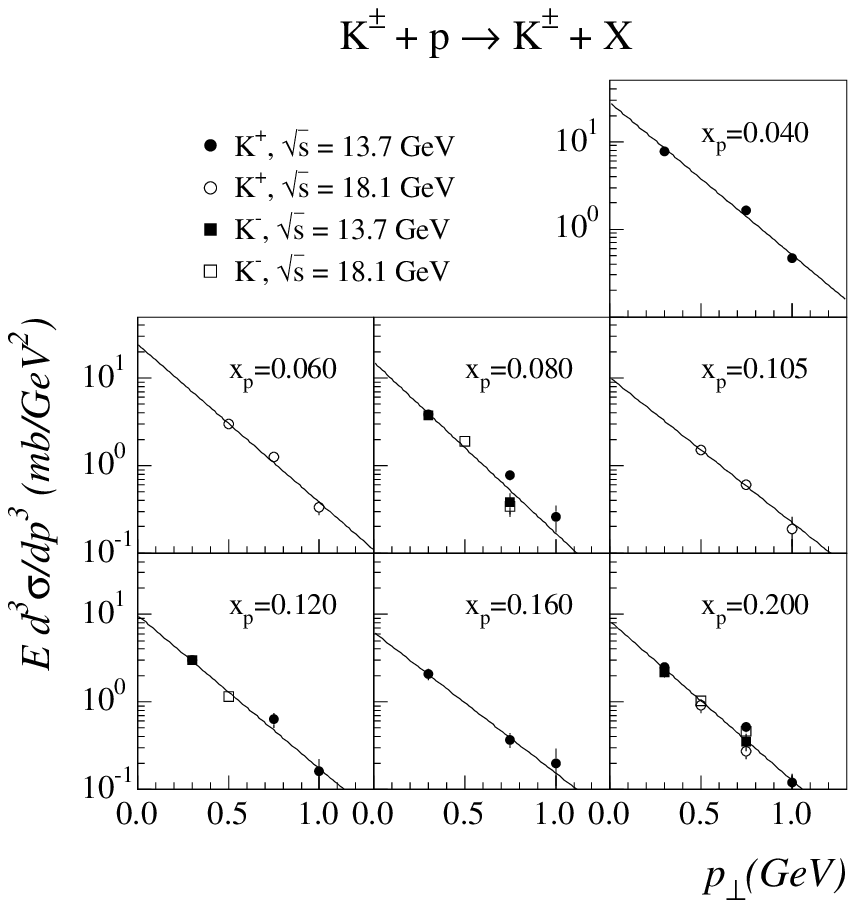} 
\caption{The invariant cross section $E\rd^3\sigma /\rd p^3 \propto
  \rd^2\sigma /\rd x_\rP\rd t$ as a function of $p_\bot \approx \sqrt{-t}$.
  Since no integrated data seems to be available, we plot different
  boxes for different values of $x_\rP$. Since the behavior in $p_\bot$
  should be independent of the value of $x_\rP$ according to
  Eq.(\ref{crosssectionmeson}), the slope parameter $a$ should be
  equal in all these boxes. This is confirmed by a fit to
  Eq.(\ref{crosssectionmeson}) which leads to values for the meson
  radii of $r_\pi^\pm =0.62 \pm 0.04$ fm, and $r_\rK^\pm =0.53 \pm 0.03$
  fm.  These values are in agreement with those obtained from form
  factor measurements (see Refs.[\ref{meson},\ref{tabelow}] for
  details).
\label{fig:meson}}
\end{figure}

We can go one step further and examine inelastic diffractive reactions
off other hadrons such as $\rK^\pm$ and $\pi^\pm$ mesons. If the proposed
picture is right, the differential cross section for such reactions
should be given by
\begin{equation}
\label{crosssectionmeson}
\frac{\rd\sigma}{\rd t} = C \cdot \exp (-2 a_{\rK,\pi} \sqrt{-t}),
\end{equation}
where $a_{\rK,\pi}=\sqrt{\frac{3}{5}}r_{\rK,\pi}$ is directly related to 
the corresponding meson radius. The comparison with the experimental data 
is shown in fig. \ref{fig:meson}.  

In summary: Systems of interacting soft gluons should be considered as
open, dynamical, complex systems far from equilibrium. The fingerprints
of SOC exist in inelastic diffractive scattering. The colorless
objects which play the dominating role in such reactions are
BTW-avalanches in the form of color-singlet gluon clusters due to SOC in
the gluon system in the proton. The differential cross section
$\rd\sigma/\rd t$ can be calculated in an optical geometrical model
taking the SOC properties of the gluon system into account and is
given by Eq.(\ref{crosssection}). The slope $a$ in
Eq.(\ref{crosssection}) is directly related to the radius of the
target hadron, which remains intact in the reaction. There is good
agreement with the experimental data. 

\section*{Acknowledgments}
I would like to thank the organizers of the conference for the interesting 
and fruitful meeting and for giving me the opportunity to present this talk.

\end{document}
